\documentclass[12pt,twoside]{article}
\usepackage{amssymb}
\usepackage{amsmath}
\usepackage[makeroom]{cancel}
\usepackage{latexsym}
\usepackage{longtable}
\usepackage{epsfig}
\usepackage{graphicx,bbm,psfrag}
\graphicspath{{images/}}

\begin{document}
\vspace*{2cm}
\begin{center}
\textbf{\Large{Domain Wall Fluctuations of the Six-Vertex Model\medskip\\ at the Ice Point}}\bigskip
 \bigskip\end{center}
 \begin{center}
\large{Michael Pr\"{a}hofer and  Herbert Spohn}\bigskip
\end{center}
 \begin{center}
Zentrum Mathematik and Physik Department, TUM,\\
Boltzmannstr. 3, 85747 Garching, Germany.\\
 \tt{praehofer@ma.tum.de, spohn@tum.de}
 \end{center}
%\maketitle
\vspace*{3cm}
\textbf{Abstract}. We report on Monte-Carlo simulations of the six-vertex model  with domain wall boundary conditions. In thermal equilibrium such boundary conditions force a fluctuating line separating
the disordered region from the perfectly ordered ones. Specifically we study the ice point at which all vertex weights are equal. With high precision the one-point fluctuations of the line are confirmed to be of order $N^\frac{1}{3}$
and governed by the Tracy-Widom distribution. Furthermore, the non-universal scaling coefficients 
are computed for a wide range of interaction strengths. 
\vspace*{1cm}
%\begin{flushright} 1.5.2023
%\end{flushright}
\newpage
%%%%%%%%%%%%%%%%%%%%%%%%%%%%%%%%%%%%%
%%%%%%%%%%%%%%%%%%%%%%%%%%%%%%%%%%%%%
\section{Introduction}\label{sec1}
\begin{figure}[!b]
  \centering
  \includegraphics[width=0.75\linewidth]{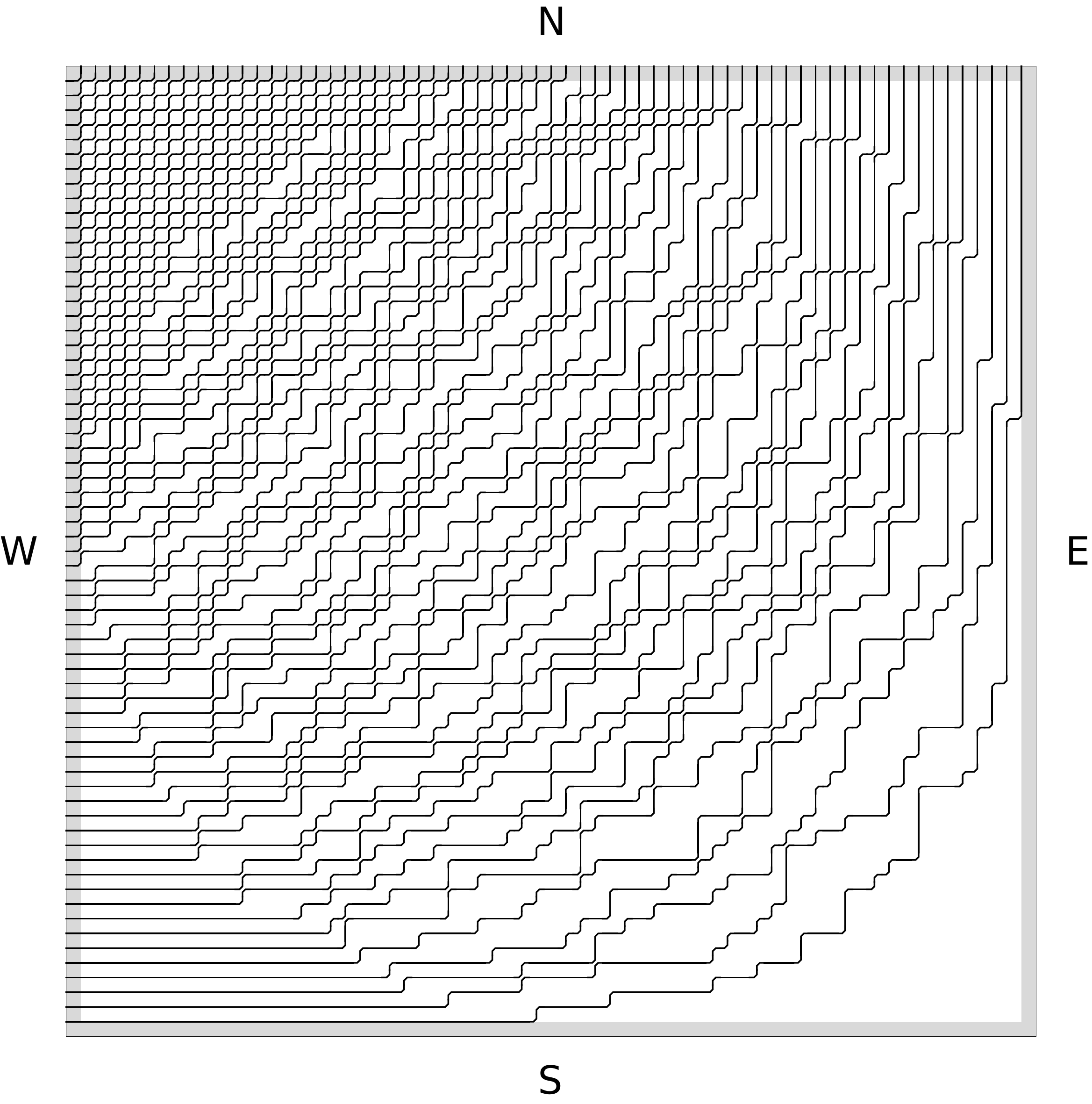}
  \caption{One realization of the level lines at $\Delta = \tfrac{1}{2}$ for $N = 64$ with DWBC. Even for such a fairly small system 
one clearly distinguishes a disordered zone located in the center from the perfectly ordered facets situated in the four corners. Our focus is the facet edge, more specifically its SE section.}
\label{fig:1}
\end{figure}

The six-vertex model is one of the most famous Bethe ansatz solvable models of Statistical Mechanics in two dimensions \cite{B87,R10}. While various physical interpretations are used, in our contribution the model will be viewed as a statistical ensemble
of possibly touching, but non-crossing up-right paths, see Figure \ref{fig:1}.  In Figure 1 we imposed domain wall boundary conditions (DWBC), which means that the lines are kept fixed at the boundary of the square. The lattice paths will be viewed as level lines of a height function,
which takes the value 0 in the South-West corner and increases by 1 whenever crossing a level line. 
%%%%%%%%%%%%%%%%%%%%%%%%%%%%%
%%%%%%%%%%%%%%%%%%%%%%%%%%%%%
Thus the height function takes 
the value 0 at the South (S) and East (E) borders, while increasing as a slope 1 staircase at the West (W) border and decreasing as a slope $-1$ staircase at the North (N) border.  In the interior of the square the height is allowed to fluctuate.
Studied will be a random surface, which is defined through the graph
of an integer-valued random height function over the basic domain $[0,N+2]^2$ increasing in the East-to-West and South-to-North directions.  

Physically one could think of a soap bubble, i.e. an elastic membrane with non-zero surface tension which is pinned at a given frame. Another interpretation is a three-dimensional crystal, whose shape is modelled by a height function satisfying the monotonicity conditions mentioned above.

We are interested in the large $N$ limit. As suggested by Figure 1, one expects 
a deterministic  macroscopic shape. It consists of an interior zone surrounded by four perfectly ordered regions.  In the interior domain the macroscopic shape is rounded with nonzero curvature. Microscopically the height is disordered with fluctuations  of order $\log(N)$, more precisely the 
height statistics are those of a massless Gaussian field with a covariance depending on the specific macroscopic location in the interior zone \cite{BR22}. The height gradients are critical as signalled by a slow decay of their correlations. On the other hand, 
in the four corners the height function is flat with no fluctuations.

In addition to the non-crossing constraint, the lines have a nearest neighbor interaction with a strength usually denoted by $\Delta$. Its precise definition will be given below.
At $\Delta = 0$, the model can be handled using free fermion techniques. In this case an explicit formula for the macroscopic shape is available \cite{KZ00,ZJ00,CKP01,FS06, ADSV16}, which indeed has four perfectly ordered facets located in the four corners. The facet edge is an ellipse reflection symmetric with respect to the diagonals and touching each one of the four bounding edges. For general $\Delta$ the information is less complete. For $-1 < \Delta < 1$ the macroscopic shape is qualitatively similar to the one with free fermion parameters. For $\Delta > 1$ the disordered region degenerates into a line with either diagonal or anti-diagonal orientation and perfect facets in the lower respectively upper triangle. For $\Delta < -1$, minimizing the surface free energy an inner facet appears. Inside this facet the level lines are purely zig-zag with distance 1. A defect costs only an energy of order one.
Hence  statistical errors can be accommodated with nonzero density, decreasing to zero as 
$\Delta \to -\infty$. The outer facet edge gets further deformed, but in essence does not notice the birth of the inner facet. The disordered zone 
has the shape of  an annulus. We refer to \cite{ZJ10} for typical line configurations at various choices of $\Delta$.

The macroscopic shape  has a universal feature known as Pokrovsky-Talapov (PT) law, which refers to macroscopic crystal shapes \cite{PT79}.
Close to the facet edge the deviation from the facet plane varies with the position $x$ relative to the facet edge as
% Along a line perpendicular to the facet edge the crystal shape has schematically the form
$s(x)=0$ for $x\leq0$ and $s(x)=\pm x^\delta$ for $x>0$, the left describing the facet and the right the rounded part. The PT law predicts in our context the value of $\delta$ to be $\delta=3/2$.
%The rounded piece joins the flat facet with a $\tfrac{3}{2}$-power law.
This implies that close to the facet the line density has a square root singularity, which in turn translates into a distance of order  $N^\frac{1}{3}$ between neighboring level lines \cite{FPS04}.  
In case of DWBC, except for the free fermion parameters, the analytical formula for the macroscopic shape is not sufficiently explicit to deduce PT. For the outer facet edge, however, a parametric representation has been accomplished \cite{CP10,CPZ10,CS16,A20}. The representation is based on exact hole probabilities, i.e. the probability that the rectangle $[n,N+2]\times [0,m]$ is free of lines. 

The focus of our contribution are fluctuations of the facet edge. Based on a random walk analogy one might expect that the
fluctuations are Gaussian with size $N^\frac{1}{2}$, but this would miss the non-crossing constraint between neighboring level lines. In fact, assuming the validity of PT,  the facet edge fluctuations are expected to be of order $N^\frac{1}{3}$. Finer details  
can be understood through the connection with stochastic growth processes. For this purpose one regards $N$ as a time parameter. Increasing the system size from $N$ to $N+1$ induces a random growth of the facet edge. 
At the free fermion point the update turns out to be governed by a Markov chain with a structure very similar to the discrete time asymmetric simple exclusion process starting from step initial conditions. With such input, it is ensured that   the
Kardar-Parisi-Zhang (KPZ) theory of growing interfaces \cite{KPZ86} applies. In particular,  as established in \cite{J00}, in the limit of large $N$ the facet edge statistics at a single reference point  is determined by the Tracy-Widom distribution of the largest eigenvalue of a $N\times N$ GUE random matrix. 

With this background one might ask whether the GUE Tracy-Widom distribution is special for the free fermion parameters or possibly a generic feature of the 
six-vertex model. Amongst experts, the latter version seems to be favored. But when pressed the actual evidence is scarce. One natural approach is to test through Monte Carlo simulations \cite{WJ05,AR05,LKV17}. However size and number of samples are not large enough for our purposes. More extensive Monte Carlo simulations are reported in \cite{LKV23}.  We try to improve
the situation through Monte-Carlo simulations of the six-vertex model at the diagonal point with $\Delta= \tfrac{1}{2}$. This specific choice of parameters is known as ice point for which all admissible height configurations have the same weight. The corresponding growth process is no longer a Markov chain. But  KPZ scaling theory is still applicable. We then follow the standard route and first 
 determine the non-universal coefficients of the KPZ scaling theory \cite{T18}. They are expressed in terms of the macroscopic edge function and the derived formulas are valid for the outer edge at any $\Delta$. Thereby we arrive at a parameter-free numerical fit of our numerical data.
 
 From the probabilistic side indirect but very strong evidence has been obtained. According to KPZ theory, relative to a coordinate system rotated by $\pi/4$, the minimum of the SE facet edge
 should be distributed according to the GOE Tracy-Widom distribution \cite{J03}. 
 For parameters at the ice point this property has been proved recently in \cite{ACJ23}.

 The ice point is also of interest in combinatorics, more specifically for random alternating sign matrices (ASM). A matrix $A_{ij}, i,j=1,...,N+1$, is ASM if 
$A_{ij}$ takes values $0,\pm 1$ and, along each row and column, the $+1$ and $-1$ entries alternate and sum up to 1. Investigated are statistical properties of ASM under the uniform distribution
\cite{BP99,B99}. For DWBC the connection is based on a mapping between ASM and vertex configurations under which the uniform distribution of ASM matrices corresponds to the ice point. Thus our results also bear on fluctuations of random ASMs. 

In the following section we define the six-vertex model and discuss its phase diagram for the case of DWBC. 
As input to the KPZ scaling theory the non-universal coefficients  are computed for the outer facet edge. We discuss Monte Carlo (MC) schemes.
The numerical simulations have been carried out for the ice point labeled as ``ASM'' and for the symmetric free fermion point tagged as ``Aztec'' for reasons explained below. Our results are displayed in Section 5, including a detailed comparison with the predictions of the theory.
%%%%%%%%%%%%%%%%%%%%%%%%%%%%%%%%%%
\begin{figure}[!t]
  \centering
  \begin{tabular}{cccccc}
    1&2&3&4&5&6\\
  \includegraphics[width=2cm]{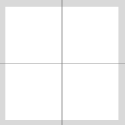}&
  \includegraphics[width=2cm]{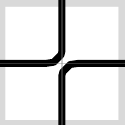}&
  \includegraphics[width=2cm]{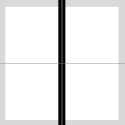}&
  \includegraphics[width=2cm]{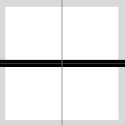}&
  \includegraphics[width=2cm]{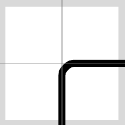}&
  \includegraphics[width=2cm]{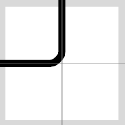}    
  \end{tabular}
  \caption{Tiles of the six-vertex model with weights $\omega_\alpha$, $\alpha = 1,...,6$.}
  \label{fig:2}
\end{figure}
%%%%%%%%%%%%%%%%%%%%%%%%%%%%%%%%%%%%%%%%%%%%%%%%%%
%%%%%%%%%%%%%%%%%%%%%%%%%%%%%%%%%%%%%%%%%%%%%%%%%%
%\newpage
\section{Domain wall boundary conditions}\label{sec2}
The most direct route to the six-vertex model is through a tiling problem. Given are six distinct tiles of side-length $1$,
%with strictly positive Boltzmann weights $\omega_\alpha$, $\alpha = 1,...,6$,
where we  follow the convention of \cite{FS06,CS16},  see Figure 2. An admissible tiling is defined by having no broken lines. 
For DWBC we consider the square, $\Lambda_N = [0,N+2]^2 \subset \mathbb{R}^2$, of side length $N+2$, with the origin at the SW corner point, compare with Figure 3. The boundary region is $[0,N+2]^2\setminus [\tfrac{1}{2},N +\tfrac{3}{2}]^2$. The West boundary has horizontal line segments placed at integers and the North boundary has vertical line segments also placed at integers. 
%%%%%%%%%%%%%%%%%%%%%%%%%%%%%%%%%%
\begin{figure}[!b]
  \centering
  \includegraphics[width=0.35\linewidth]{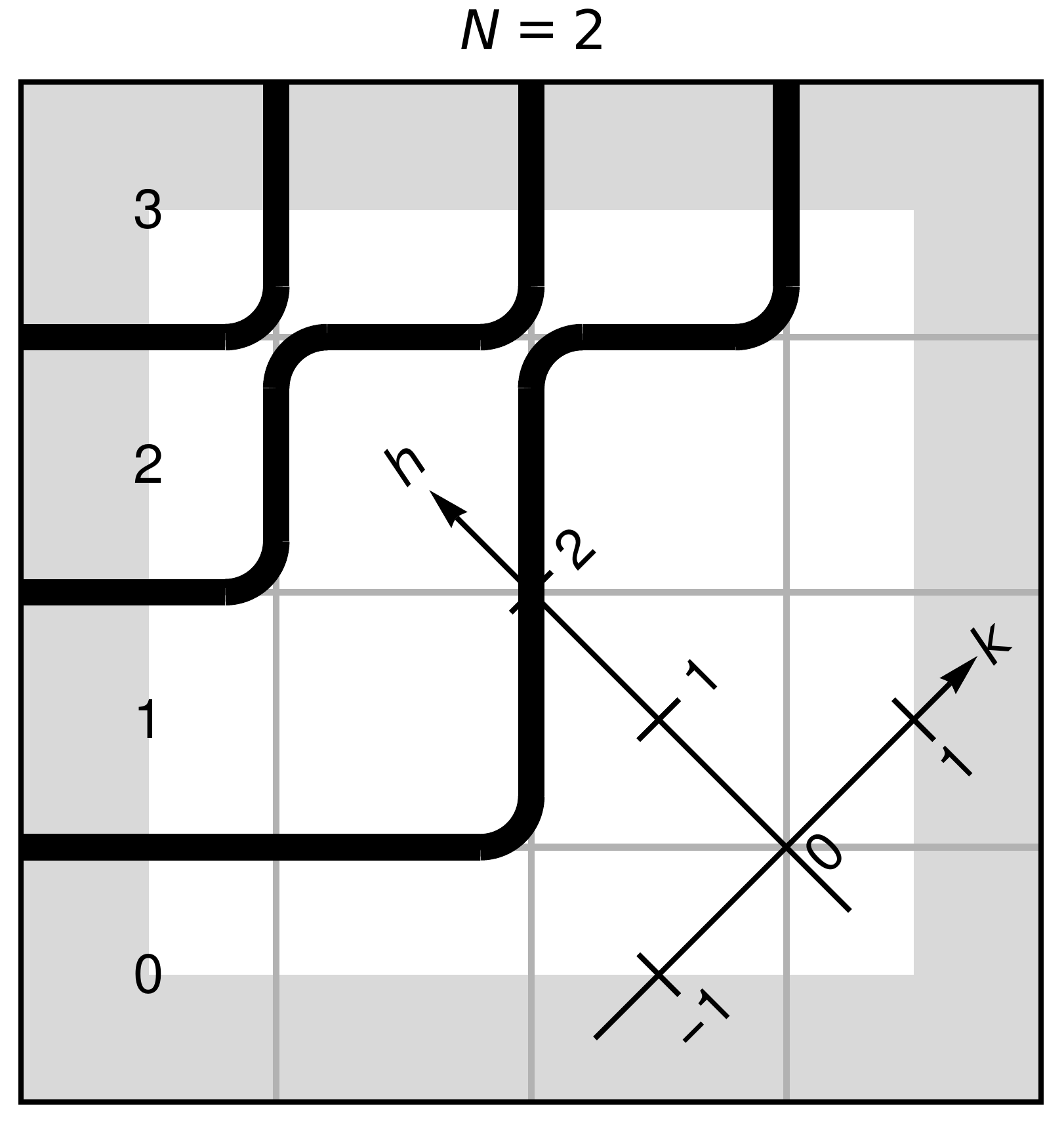}
  \caption{One realization of the level lines for $N = 2$ for prescribed boundary tiles. For better visualization the tiles  themselves are not shown.}
\label{fig:3}
\end{figure}
%%%%%%%%%%%%%%%%%%%%%%%%%%%%%%%%%%
The square $[\tfrac{1}{2},N +\tfrac{3}{2}]^2$ is tiled under the constraint that the thereby generated lines end only at the West and North edge. The six tiles have Boltzmann weights $\omega_\alpha>0$, $\alpha = 1,...,6$.
The weight of an admissible tiling is given by the product
\begin{equation}\label{1.1}
\prod_{(i,j)} \omega_{\alpha(i,j)}
\end{equation}
where $\alpha(i,j)$ is the index of the tile centered at $(i,j)$ and $i,j = 1,...,N+1$.
An admissible tiling
corresponds to a collection of up-right lattice paths, whose possible turns are only at the grid points of $\mathbb{Z}^2$.
The paths may touch, but they never cross.  
These paths are the level lines of a height function $\phi_N(x_1,x_2)$, $0 \leq x_1,x_2\leq N+2$, which is constant on each unit square,
increases by one at each NW oriented crossing of a level line, and is normalized as $\phi_N(0,N+2)=N+1$.

The conventional parametrization of the Boltzmann weights is given by 
   \begin{equation}\label{1.5}
 \omega_1= ae^{H+V},\quad \omega_2=ae^{-H-V},\quad \omega_3 = be^{H-V},\quad \omega_4 = be^{-H+V}, \quad \omega_5 = c, \quad \omega_6 =c.
\end{equation} 
The free energy depends on $a,b,c$ only through $a/c$ and $b/c$, which defines the interaction  strength
\begin{equation}\label{1.4}
\Delta = \frac{a^2 +b^2 - c^2}{2ab}.
\end{equation}
$H,V$ control the slopes of $\phi_N$. For DWBC it suffices to set $H=0=V$, which leaves 
$a/c,b/c$ as free parameters. Symmetric case means $a = b$. The free fermion parameters form the quarter circle defined by $\Delta = 0$, see Figure 4.
Ice point refers to $a=1$, $b=1$, $c=1$, which implies a uniform distribution
over all admissible line configurations. In the following we use the
acronym ASM for the ice point. The bijection from 
%%%%%%%%%%%%%%%%%%%%%%%%%%%%%%%%%%%%%%
\begin{figure}
  \centering
  \includegraphics[width=0.55\linewidth]{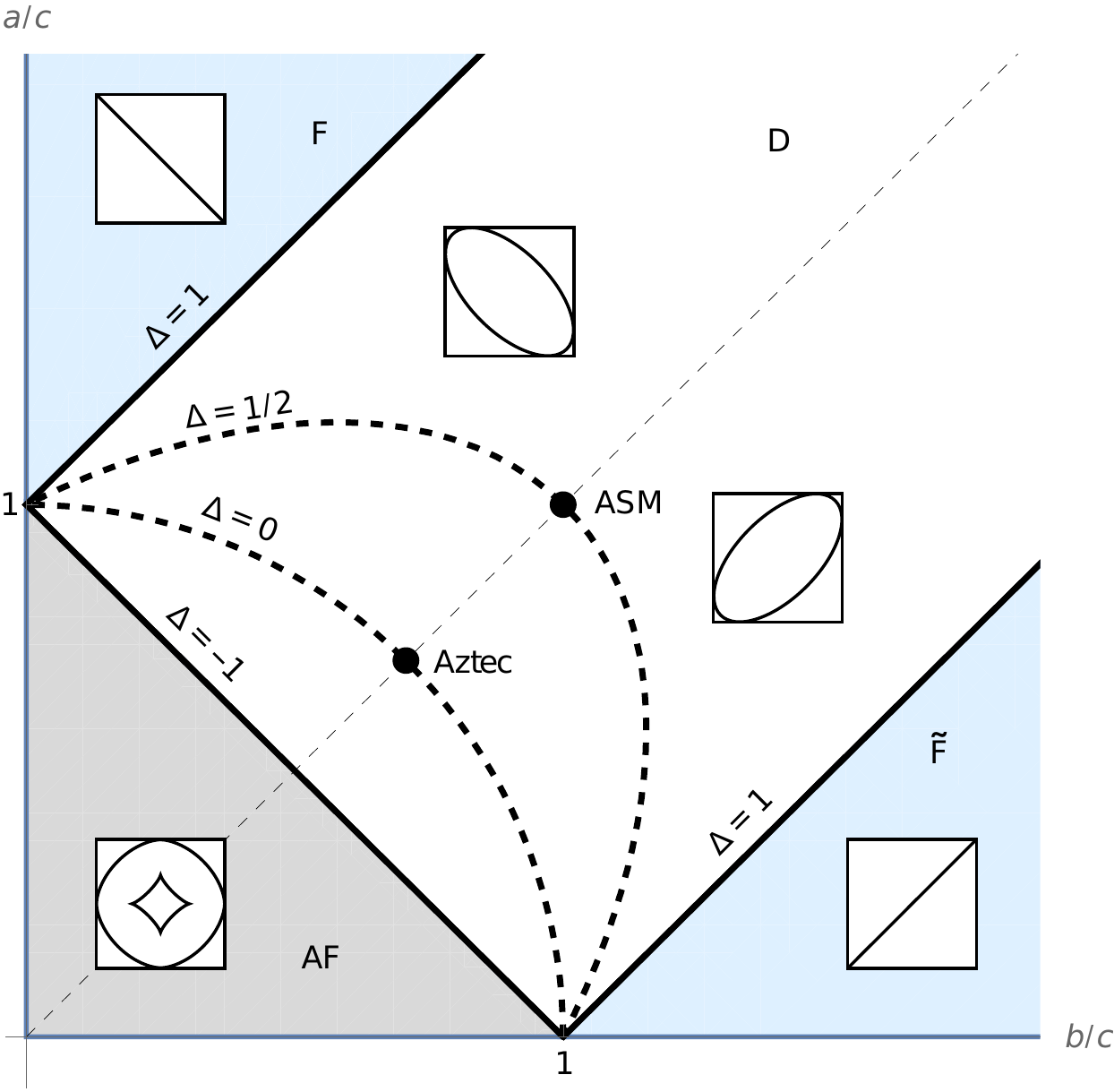}
  \caption{Phase diagram of the six-vertex model with DWBC.}
  \label{fig:4}
\end{figure}
%%%%%%%%%%%%%%%%%%%%%%%%%%%%%%%%%%%%
admissible DWBC configurations to $(N+1)\times(N+1)$ matrices is given by
\begin{align} 
  A_{ij}=
  \begin{cases}
    -1&\mbox{for $\alpha(i,j)=5$},\\
    +1&\mbox{for $\alpha(i,j)=6$},\\
    0&\mbox{otherwise.}
  \end{cases}
\end{align}
Using DWBC it follows that the so defined matrix is ASM.

The model at the symmetric free fermion point, $a=b=1$, $c=\sqrt2$, will be labelled as ``Aztec'' for the following reason: Due to DWBC %the domain wall boundary conditions
one can equivalently assign the weights $\omega_1=\cdots=\omega_4=\omega_6=1$, $\omega_5=2$, corresponding to the so called $2$-enumeration of ASMs, where each $-1$ in a given ASM has weight $2$.
As explained in \cite{FS06} in a more general context, there is a natural one-to-one correspondence between domino tilings of an Aztec diamond with size $N$ and $2$-enumerated ASMs of size $N+1$ (to account for the 2-enumeration one introduces a $7$th tile of type, say, $5'$, which is a twin of tile $5$. Domino tilings can then be translated into vertex configurations via a local rule). 
%Introducing a $7$th tile of type $5'$, which can be freely interchanged with tile $5$, the resulting seven vertex model with DWBC is in one-to-one correspondence with uniform domino tilings of a size $N$ Aztec diamond.
In this case the macroscopic facet edges form a circle which has been termed arctic circle, since it separates the four frozen zones at the corners from the disordered zone in the interior.

For general weights, one expects that there is a deterministic macroscopic limiting shape function $\phi_\mathrm{ma}$
such that
\begin{equation}\label{1.6}
 \lim_{N\to \infty} N^{-1}\phi_N (\lfloor Ny_1 \rfloor,\lfloor Ny_2 \rfloor) = \phi_\mathrm{ma}(y_1,y_2),\quad 0 \leq y_1,y_2\leq1,
\end{equation}
$\lfloor \cdot \rfloor$ denoting integer part. In Figure 4 we display the phase diagram for $\phi_\mathrm{ma}$.
For the domain D (disordered), 
$\phi_\mathrm{ma}$ has perfect facets in the four corners of $[0,1]^2$, clockwise with tiles of type 1,4,2,3, compare with  Figure 1 and 2,
and interpolates smoothly with a non-zero curvature in-between. 
 In the domain AF (anti-ferroelectric) in addition there appears a central facet, isolated from the outer facets, which has a pattern consisting
 of alternating tiles of type 5,6 with Poisson like random errors.  
In the domains F and $\tilde{\mathrm{F}}$ (both ferroelectric) the interior zone degenerates to a line, either diagonal or anti-diagonal. The facets consist either of type 1,2 (diagonal interface) or type 3,4 (anti-diagonal interface). 

For finite $N$ there is the well-known Izergin-Korepin formula \cite{ICK92} for the free energy. Its asymptotics, including subleading terms, is scattered in various contributions and summarized in \cite{BL13}. Some qualitative features of the phase diagram can be recovered. The  features for interest to us concern the local free energy, which is beyond the  Izergin-Korepin formula.
%%%%%%%%%%%%%%%%%%%%%%%%%%%%%%%%%%%%%%%%%%%%%%%%%%%
%%%%%%%%%%%%%%%%%%%%%%%%%%%%%%%%%%%%%%%%%%%%%%
\section{Scaling theory}\label{sec3}
We turn to the  finer structure of the facet edge in case of DWBC and parameters in the domain D. 
By symmetry, it suffices to consider only a quarter section of the edge, for which we choose the  SE facet, i.e. $\{(x_1,x_2)\in[0,N{+}2]^2\,|\,\phi_N(x_1,x_2) = 0\}$.
It will be convenient to view its NW boundary as the graph of a function $h_N$ in a coordinate system centered at $(N+1,1)$ and rotated by $\pi/4$, compare with Figure 3.
We refer to $h_N$ restricted to $[-N,...,N]$ as the edge function. One always has $h_N(-N) = N =h_N(N)$. Furthermore $h_N$ is piecewise linear between integers and takes 
integer values on $\{-N,\dots,N\}$ with increments being $\pm 1$.
% Its NW boundary is given through the graph of a function $h_N$.  For convenience we refer to  $h_N$ as edge function.  It will be convenient to view
%$h_N$ in a coordinate system rotated by $\pi/4$ and centered at $(N,1)$.
%changes MP
%$h_N$ is piecewise linear between integers in $[-N,...,N]$ and takes 
%integer values on $\{-N,\dots,N\}$ with increments being $\pm 1$, and $h_N(-N) = N =h_N(N)$.
 For example, the edge function $h_N$ corresponding to Figure 3, where $N=2$, is specified by the values
\begin{align*}
  \begin{array}{c||c|c|c|c|c}
    k&-2&-1&0&1&2\\
    \hline
    h_N(k)&2&1&2&3&2
  \end{array}
\end{align*}
Attaining a macroscopic shape as in \eqref{1.6} strongly indicates the existence of a limiting edge  function $h_\mathrm{ma}$ such that
\begin{equation}\label{3.1}
 \lim_{N \to \infty} N^{-1}h_N\big(\lfloor N x\rfloor\big) = h_\mathrm{ma}(x),\quad |x| \leq 1.
 \end{equation}
 A parametric representation of $h_\mathrm{ma}$ is computed in \cite{CP10,CPZ10}, see also the recent contributions \cite{CS16,A20}. 
  
 Assuming that the edge fluctuations are governed by the KPZ universality class, one expects that  
\begin{equation}\label{3.2}
h_N\big(\lfloor Nx\rfloor\big) \simeq h_\mathrm{ma}(x) N -
(\Gamma(x)N)^\frac{1}{3}\xi_\mathrm{TW},\quad |x| < \tfrac12,
\end{equation} 
to leading order in $N$ \cite{FPS04}. Here $\xi_\mathrm{TW}$ is distributed according to the Tracy-Widom distribution from GUE random matrix theory with distribution function $F_2(s)=\mathrm{Prob}(\xi_\mathrm{TW}\leq s)$ and $\Gamma(x)$ is a model dependent 
parameter. The Tracy-Widom distribution has a negative mean. Therefore Eq.~\eqref{3.2} indicates 
that the actual first level line is slightly above the macroscopic edge profile. One knows that for $|x|=\frac12$ the limit \eqref{3.2} fails  \cite {JN06,G14} and  convergence non-uniform in $x$ is the rule. The most rapid convergence is expected to be close to $x = 0$.   The KPZ scaling theory relates $\Gamma(x)$ to more directly accessible properties of the model
\cite{KMH94}. For DWBC this amounts to linking $\Gamma(x)$ with $h_\mathrm{ma}(x)$. In more complicated models such input might not be available and 
has then to be extracted from the numerical data (or from an actual experiment \cite{T12}).

The KPZ scaling theory starts from a growth model, which on the macroscopic scale has a growth velocity $v(\partial_x h)$ depending on the local slope 
$\partial_x h$. The shape of the growing cluster is  hence governed by
\begin{equation}\label{3.3}
\partial_t h(x,t) = v\big(\partial_x h(x,t)\big) .
\end{equation} 
For the six-vertex model with DWBC the size $N$ plays the role of the time parameter. We do not attempt to write down a stochastic dynamics
linking $N$ to $N+1$. But one can still expect the validity of \eqref{3.3} for a suitable choice of $v$. If \eqref{3.3} holds,
then in our context $h_\mathrm{ma}$ must be determined through a self-similar solution of the form $h(x,t) = t h_\mathrm{ma}(t^{-1}x)$ and
one concludes that 
$h_\mathrm{ma}(x)$ is the Legendre transform of $v(u)$, $u = h_\mathrm{ma}'$. In particular
 \begin{equation}\label{3.4}
h_\mathrm{ma}''(x)v''\big(u(x)\big) = -1.
\end{equation}
In the next few lines  we suppress the $x$-dependence for conciseness.

The KPZ scaling theory \cite{T18} asserts that 
\begin{equation}\label{3.5}
  % \Gamma = (\tfrac{1}{2} |v''|A^2)^\frac{1}{3} .
  \Gamma = \tfrac{1}{2} |v''|A^2 .
\end{equation}
The coefficient $A$ is determined by the local roughness of the interface. More precisely, for large $N$, $j \mapsto h_N\big(\lfloor Nx\rfloor +j \big)$
looks like a random walk with drift, implying the variance 
\begin{equation}\label{3.6}
\mathrm{Var}\big( h_N\big(\lfloor Nx\rfloor +j\big)- h_N\big(\lfloor Nx\rfloor\big)\big) = A|j|.
\end{equation}
Note that here one requires $1 \ll  |j| \ll N^\frac{2}{3}$, the latter being the scale of mesoscopic fluctuations.

It remains to compute $A$. For large $N$ the edge  $h_N$ is separated from the neighboring level line by $N^\frac{1}{3}$. Thus, for the computation of $A$,
contacts between the two lines can be ignored  and it suffices to only consider a single up-right lattice path. The number of tiles 1 and 2 is not modified. The up- and right-unit segments lines have weights
$be^\lambda$ and $be^{-\lambda} $, where $c=1$ without loss of generality
% for simplicity
and we introduced the parameter $\lambda$ to control the bias.
% resp., where $\lambda$ controls the bias and $c=1$ for notational simplicity.
We use the diagonal transfer matrix, consistent with the frame introduced 
 for $h_N$. The two-step transfer matrix $T$ is obtained as
 \begin{eqnarray}\label{3.7}
&&T_{i,i-1} = be^{-\lambda},\quad  T_{i,i}= 1,\quad T_{i,i+1}=be^{\lambda},\quad T_{i,i+2} = b^2e^{2\lambda},\nonumber\\
&&T_{i+1,i-1} = b^2e^{-2\lambda},\quad T_{i+1,i}= be^{-\lambda} ,\quad T_{i+1,i+1}= 1,\quad T_{i+1,i+2}= be^{\lambda},
\end{eqnarray} 
for $i$ even, with all other matrix elements vanishing.
$T$ is a two-periodic Toeplitz matrix. Summing over all $2n$-step walks, $(X_i)_{i=0,\dots,2n}$ with $\sigma_i:=X_{i}-X_{i-1}\in\{\pm1\}$, $i=1,\dots,2n$, starting with $X_0=0$ or $X_0=1$ and allowing for arbitrary end points, yields the partition function
\begin{align}
  Z_{2n}(\lambda)
  &=\sum\limits_{\sigma\in\{\pm1\}^{2n}}\prod\limits_{i=1}^{2n-1}b^{(\sigma_i\sigma_{i+1}+1)/2} \mathrm{e}^{\lambda(\sigma_i+\sigma_{i+1})/2}\\
  &=\sum\limits_{j\in\mathbb{Z}}\big((T^n)_{0j}+(T^n)_{1j}\big)
  =\langle\psi,L(\lambda)^n\psi\rangle
\end{align}
with $\psi=(1,1)$. Here the $2\times 2$ matrix $L(\lambda)$ is given by
\begin{equation}\label{3.9}
L(\lambda)=\begin{pmatrix}
1+ b^2e^{2\lambda}&2b \cosh \lambda\\
2b \cosh \lambda & 1+ b^2e^{-2\lambda}.
\end{pmatrix}.
\end{equation}

For large $n$ the partition function is dominated by 
 the largest eigenvalue of $L(\lambda)$, which is determined to  
 \begin{equation}\label{3.10}
E(\lambda) = 1 + b^2\cosh 2 \lambda + \sqrt{ (1+b^2 \cosh 2 \lambda)^2 - (1-b^2)^2}.
\end{equation} 
Hence 
\begin{equation}\label{3.11}
Z_{2n}(\lambda) \simeq E(\lambda)^n.
\end{equation} 
Asymptotically the mean of the random walk is given by $(Z_{2n}'/Z_{2n})\simeq n(E'/E) = 2nu$, where $u$ is equated with $h'_\mathrm{ma}$.  By definition the variance  $(Z_{2n}'/Z_{2n})'\simeq n(E'/E)' = 2nA$. Somewhat unexpectedly, the asymmetry parameter
$\lambda$ can be eliminated explicitly from the two equations
\begin{align}
  u&=\frac{b(\mathrm{e}^{2\lambda}-1)}{\sqrt{b^2(\mathrm{e}^{2\lambda}-1)^2+4\mathrm{e}^{2\lambda}}},\qquad
     A=\frac{4b\mathrm{e}^{2\lambda}(\mathrm{e}^{2\lambda}+1)}{\big(b^2(\mathrm{e}^{2\lambda}-1)^2+4\mathrm{e}^{2\lambda}\big)^{3/2}}.
\end{align}
Reintroducing $c$ the result reads
\begin{equation}\label{3.12}
A = \big(1-u^2\big) \sqrt{u^2(1-(b/c)^2) + (b/c)^2} .
\end{equation}
Thus, according to \eqref{3.4}, \eqref{3.5}, and \eqref{3.12}, the scale coefficient $\Gamma(x)$ is determined through $h'_\mathrm{ma}(x) = u(x)$ and $h''_\mathrm{ma}(x)$ as  $\Gamma(x)=\frac12A(x)^2v''(u(x))=-\frac12A(x)^2/h_{\mathrm{ma}}''(x)$, $b/c$ indicating the parameter of the six vertex model.
 
Since there is no restriction on $b/c$, our argument remains valid for the outer facet for all $\Delta <1$.
%The parameter $u$ depends on $x$ through $h'_{\mathrm{ma}}(x)=u(x)$ and so does $\Gamma$ through $\Gamma(x)=\frac12A(x)^2v''(u(x))=-\frac12A(x)^2/h_{\mathrm{ma}}''(x)$.
The classic result of Colomo and Pronko \cite{CP10}
yields $h_{\mathrm{ma}}$ explicitly for $\Delta=\frac12$ and $\Delta=0$. Our result \eqref{3.12} allows quantitative prediction of the scaling form \eqref{3.2} along each direction $x$ with $|x|<\frac12$ to be compared with numerical results, see \eqref{aztecscaling}, \eqref{asmscaling} in Section 5.

In principle the KPZ scaling theory should also apply to the inner facet in the case $\Delta < -1$. However the required input is not yet available. Firstly the position of the macroscopic facet edge is not known sufficiently explicit. Furthermore, since the facet has Poisson type defects the microscopic facet edge is only fuzzily defined. To obtain the coefficient $A$ one had to rely on a more sophisticated reasoning.

%Since there is no restriction on $b/c$, our argument remains valid for the outer facet for all $\Delta <1$. In principle the KPZ scaling theory also applies to the inner facet, when $\Delta < -1$. However the required input is not yet available. Firstly one would have to compute the macroscopic facet edge. In addition, since regular order is perturbed by Poisson type noise, to obtain the $A$ coefficient has to rely on a more sophisticated reasoning.

%Such a relation holds for all $\Delta < 1$. Of course, in the AF domain the argument refers only to the outer facet. The edge of the central facet has no sharp
%microscopic definition and even if there would be a practical rule, the computation of $A$ cannot be based on only a single lattice path.
%%%%%%%%%%%%%%%%%%%%%%%%%%%%%%%%%%%%%%%%%%%%%%%
%%%%%%%%%%%%%%%%%%%%%%%%%%%%%%%%%%%%%%%%%%%%%%%%
\section{Monte Carlo simulations}\label{sec4}
We construct a Markov chain in such a way that its unique stationary measure is the normalized Gibbs distribution \eqref{1.1} of the six-vertex model with DWBC. First we  explain the standard detailed balance approach,  discussing further options at the end of the section.
For the Monte Carlo scheme used here, it is convenient to view the volume under the height function $\phi_N$ as made up of unit cubes stacked on top of each other. An allowed MC move corresponds to adding or removing a single cube in such a way, that the six-vertex constraint is maintained. The MC moves are conveniently encoded by introducing plaquettes, each consisting of four tiles arranged as a square, see Figure 5.
%%%%%%%%%%%%%%%%%%%%%%%%%%%%%%%%%%%%% 
\begin{figure}[!b]
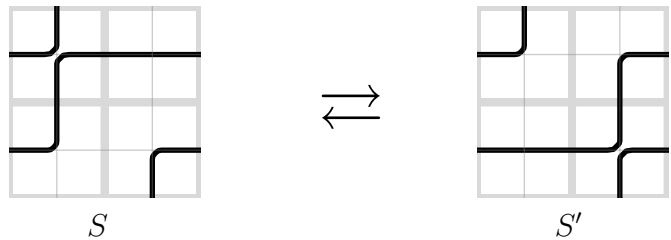

  \begin{center}
    \mbox{}\vspace{4ex}\\
  \begin{minipage}{3.3cm}
  \includegraphics{6vertexa2}\!\!
  \includegraphics{6vertexb1}\\[-.18ex]
  \includegraphics{6vertexc2}\!\!
  \includegraphics{6vertexc1}\\
  \mbox{}\hspace{2.5em}$S$
\end{minipage}{\huge\quad$\rightleftarrows$}\qquad\quad
  \begin{minipage}{3.3cm}
  \includegraphics{6vertexc2}\!\!
  \includegraphics{6vertexc1}\\[-.18ex]
  \includegraphics{6vertexb1}\!\!
  \includegraphics{6vertexa2}\\
  \mbox{}\hspace{2.5em}$S'$
\end{minipage}\vspace{-2ex}
\end{center}
  \caption{An example of a plaquette $S$ with update $S'$.}
  \label{fig:5}
\end{figure}
%%%%%%%%%%%%%%%%%%%%%%%%%%%%%%%
There are $6^4$ distinct  plaquette states. The total number of admissible plaquette states turns out to be $82$.
%Only transitions between plaquette states which leave the outgoing lines unchanged are allowed.  
%Out of these only $32$ can undergo a transition which leaves the outgoing lines unchanged. To enumerate all possible transitions they can be further grouped into $16$ pairs of plaquette states with identical outward lines. 
Out of these only $32$ are allowed to undergo a transition. To enumerate all possible transitions one groups them into 16 pairs of plaquette states having identical outward lines.
Let us denote by $(S,S')$ such a pair of plaquette states, such that $S'$ arises from $S$ by adding a unit cube in the height function picture as in Figure 5.
%The transition probability for a jump from $S$ to $S'$ is then denoted by $p_+(S \rightarrow S')$ and for the corresponding jump from $S'$ to $S$ by $p_-(S' \rightarrow S)$. Of course, $p_+(S \rightarrow S) + p_+(S \rightarrow S')= 1$ and $p_-(S' \rightarrow S') + p_-(S' \rightarrow S) = 1$. 
The probability to jump from $S$ to $S'$ is denoted by $p_+(S \rightarrow S')$. With probability $p_+(S \rightarrow S)=1-p_+(S \rightarrow S')$ there is no jump. Correspondingly we denote the jump from $S'$ to $S$ by $p_-(S' \rightarrow S)$ and the probability of no jump is $p_-(S' \rightarrow S')=1-p_-(S' \rightarrow S)$.
The condition of detailed balance is imposed as
\begin{equation}\label{4.1}
w(S) p_+(S \rightarrow S') = w(S')p_-(S' \rightarrow S),
\end{equation}
where $w(S)$ is the weight of the plaquette state $S$ as in \eqref{1.1}.  To write down more explicitly the ratio $ w(S')/w(S)$ we introduce four binary variables $\eta_\mathrm{NE},
\eta_\mathrm{SW},\eta_\mathrm{NW},\eta_\mathrm{SE}$ taking values $0,1$ and referring to the respective tile of plaquette $S$,
%. The correspondence is assigned according to the following table
\begin{align}\label{4.2}
  \eta_\mathrm{NE} &=
  \begin{cases}
    0&\mbox{for tile 4 at NE},\\
    1&\mbox{for tile 6 at NE},
  \end{cases}\qquad
  \eta_\mathrm{SW} =
  \begin{cases}
    0&\mbox{for tile 3 at SW},\\
    1&\mbox{for tile 6 at SW},
  \end{cases}\nonumber\\
  \eta_\mathrm{NW} &=
  \begin{cases}
    0&\mbox{for tile 2 at NW},\\
    1&\mbox{for tile 5 at NW},
  \end{cases}\qquad
  \eta_\mathrm{SE} =
  \begin{cases}
    0&\mbox{for tile 1 at SE},\\
    1&\mbox{for tile 5 at SE}.
  \end{cases}
\end{align}
%\begin{eqnarray}\label{4.2}
%\eta_\mathrm{NE} = 0 \hspace{6pt}\Rightarrow\hspace{6pt}\mathrm{tile}\,\, 4 \qquad &&\eta_\mathrm{NE} = 1 \hspace{6pt}\Rightarrow\hspace{6pt}\mathrm{tile}\,\, 6\nonumber\\
%\eta_\mathrm{SW} = 0 \hspace{6pt}\Rightarrow\hspace{6pt}\mathrm{tile}\,\, 3 \qquad &&\eta_\mathrm{SW} = 1 \hspace{6pt}\Rightarrow\hspace{6pt}\mathrm{tile}\,\, 6\nonumber\\
%\eta_\mathrm{NW} = 0 \hspace{6pt}\Rightarrow\hspace{6pt}\mathrm{tile}\,\, 2 \qquad &&\eta_\mathrm{NW} = 1 \hspace{6pt}\Rightarrow\hspace{6pt}\mathrm{tile}\,\, 5\nonumber\\
%\eta_\mathrm{SE} = 0 \hspace{6pt}\Rightarrow\hspace{6pt}\mathrm{tile}\,\, 1 \qquad &&\,\,\eta_\mathrm{SE} = 1 \hspace{6pt}\Rightarrow\hspace{6pt}\mathrm{tile}\,\, 5.\nonumber
%\end{eqnarray}
%Then with this convention,
Then, with these conventions (recall the weights $a,a,b,b,c,c$ for tiles $1$ through $6$, respectively),
\begin{equation}\label{4.3}
\frac{w(S')}{w(S)} = \Big(\frac{a}c\Big)^{2(\eta_\mathrm{NW}+\eta_\mathrm{SE} - 1)} \Big(\frac{b}c\Big)^{2(\eta_\mathrm{NE}+\eta_\mathrm{SW} - 1)}.
\end{equation}
%If the plaquette touches the boundary of $[\tfrac12,N+\tfrac32]^2$, the number of admissible transitions is even further reduced.
The detailed balance condition determines only the ratio between two corresponding transition probabilities. For parallel update schemes a conventional choice is 
\begin{equation}\label{4.4}
 p_+(S \rightarrow S') = \frac{w(S')}{w(S) + w(S')}.
\end{equation} 
  At the ASM point $ w(S')/w(S) = 1$ and all transition probabilities equal $\tfrac{1}{2}$.

 For the actual simulation we note that disjoint plaquettes can be updated in parallel and independently. Their corresponding transition probabilities have to be multiplied. There are four distinct ways to group all tiles into disjoint plaquettes. Hence  
  a complete Monte Carlo step consists of four consecutive sub-routines with parallel update of the distinct plaquette coverings. In the case of ASM  this procedure can be contracted to  two half-steps. All configurations reachable during a half-step have the same transition probability. Denoting by $k$ the number of flippable plaquettes in a half-step, one of these $2^k$ configurations can be chosen by consuming only $k$ random bits from a bitwise reliable random number generator. 
For such parallel MC update scheme any other choice of parameters complicates the algorithm and can lead to bottlenecks, i.e.~very small transition rates, which would slow down convergence. 

A generic difficulty of Markov Chain Monte Carlo simulations is to ensure that the chain equilibrates. In the community close to probability theory a popular choice is
the coupling from the past algorithm \cite{PW98}, which ensures perfect equilibration.
Early coupling from the past simulations of the DWBC six-vertex model can be found in \cite{W12}. But the so obtained number of samples is too small for a reliable check on KPZ universality.

In our simulations at the ASM point we also use the coupling from the past algorithm. The height functions, $\phi_N$, carry a natural partial ordering with a maximal (largest) and minimal (smallest) element. 
First one has to check that the stochastic dynamics respects this  partial order. With the choice \eqref{4.4}, such monotonicity is preserved provided
$a\leq c$, $b\leq c$ as can be checked by walking through all 16 possible local transitions $S\leftrightarrow S'$ \cite{ZJ18}. Particularly monotonicity holds for ASM. 
Given such input one evolves the maximal and minimal height configuration under the same space-time noise  from time $-T$, $T>0$, to time $0$. If at time $0$ both height functions agree, one has obtained a valid sample of the equilibrium distribution. If the height functions disagree,
one has to rerun, starting for example from time $-2T$, thereby retaining the noise during $[-T,0]$ from the prior run.  If there is still disagreement the procedure has to be repeated. The prescribed time $T$ has to be chosen with care. If too large, the computing time after coalescence is wasted. If too small, one has to restart the simulation leading to a considerable overhead in computing time.
To make a reliable guess, we use the statistics of the time of coalescense for moderate system sizes by following the volume difference between the two extremal configurations. Empirically, when doubling the
system size the number of steps for coalescence has to be increased by
a factor of $4.5$ approximately. For example, it takes roughly
$10^6$ Monte-Carlo steps (consisting of two half steps for the even
and odd sublattice each) for system size $N=510$ to have coalescence
with probability $>99.999\%$ rendering a negligible overhead for extending and
repeating an almost identical simulation.

There are other schemes,  which we have not tried. One option is called Gibbsian resampling. One observes that conditioned on its two neighboring level lines, a  given level line has an explicit Gibbsian distribution. Thus one sequentially equilibrates randomly chosen level lines with respect to a frozen background. Superficially similar approaches are directed-loop Monte Carlo \cite{SZ04}, loop-cluster update \cite{WJ05} and a full lattice multi-cluster algorithm \cite{KL17}.
%%%%%%%%%%%%%%%%%%%%%%%%%%%%%%%%%%%%%%%%%%%%%%%%%%%%%%%%%%%%%%%%%%%%%%%%%%%%%%%%%%%%%%%%%%%%%%
\section{Asymptotics for the ASM facet}\label{sec5}
%%%%%%%%%%%%%%%%%%%%%%%%%%%%%%%%%%
\begin{figure}[!t]
  \centering\mbox{}\\[3ex]
  \includegraphics[width=1.\linewidth]{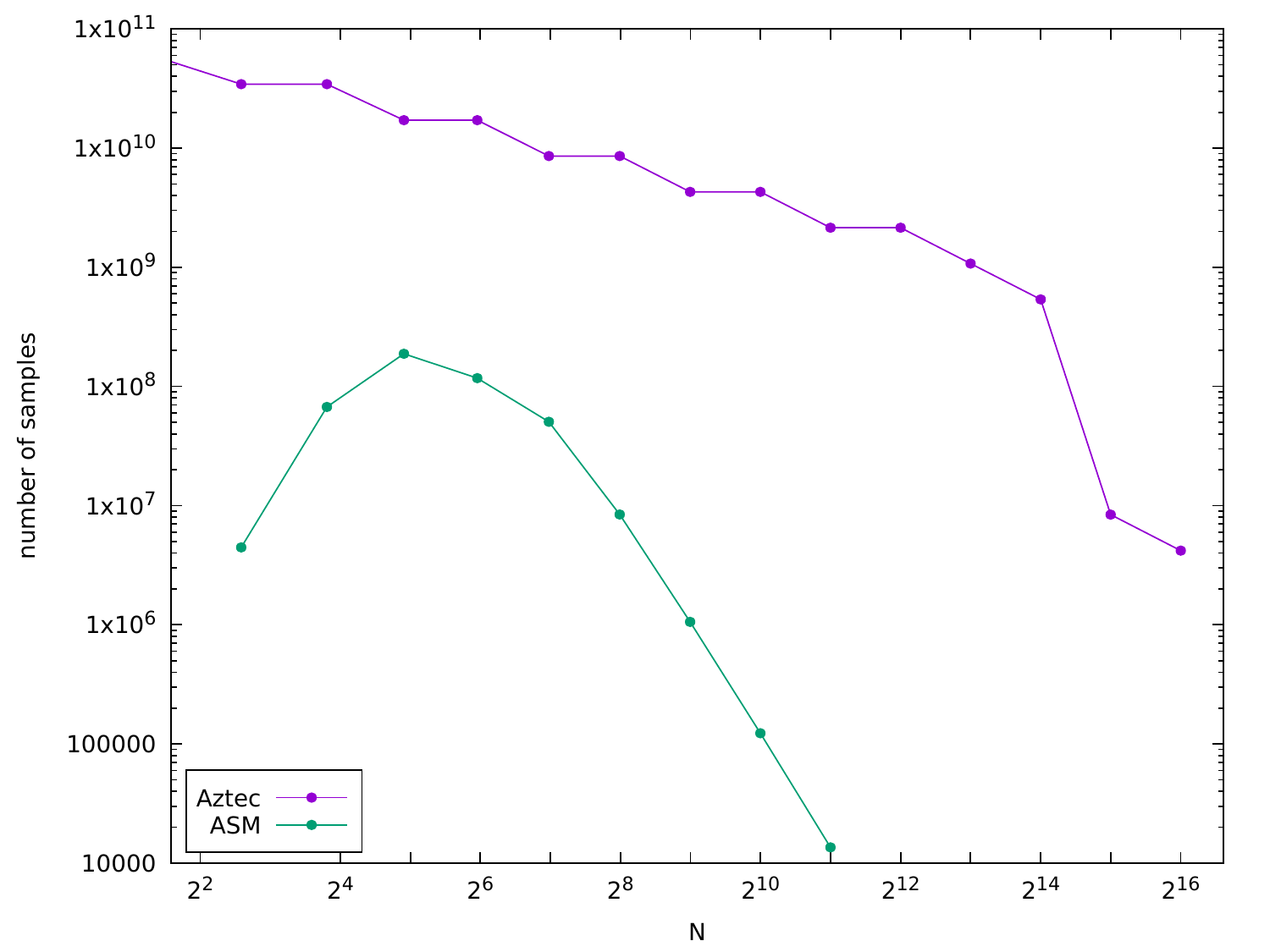}
  \caption{ASM-Aztec number of samples vs size.\\[7ex]\mbox{}}
  \label{fig:6}
\end{figure}
%%%%%%%%%%%%%%%%%%%%%%%%%%%%%%%%%
We generated samples for system sizes
$N=2,6,14,30,62,126,254,510,1022,2046$ at the ice point. Our
multi-spin implementation with 64 bit words produces 
64 independent height functions in a single run. We recorded the shape of the disordered region
for all four corners. For large $N$ these samples are essentially independent. In our 
implementation one run consumed roughly $(N/64)^4$ seconds
on a single core (as of 2018) leading to 256 almost independent corner
samples (Mersenne twister was used as pseudorandom number generator).

In our numerical data analysis, for comparison we also display results
for the free fermion point  
$a^2= b^2 = \frac{1}{2}$, which allows for well-known simplifications.
One first maps the DWBC six-vertex model to the Aztec diamond,  see
for example the  
discussion in \cite{FS06}, which is a dimer tiling of a square rotated by $\pi/4$
and width 2 at the corners. The shuffling algorithm ensures that the
steady state at size $N$ is transformed to the steady state at size
$N+1$. Secondly under the shuffling algorithm the  
microscopic facet edge evolves under its own MC dynamics known also as
a particular corner growth model, which can be viewed as a  
discrete time version of TASEP \cite{J00}. Thus, instead of simulating a
two-dimensional system, one can directly update
$h_N$ from $h_{N-1}$, with an effort only proportional to $N$. Hence compared to the ASM model much larger system sizes can be reached for Aztec.
Sample sizes for the corner growth model are $N=2^k$, $k=2,\dots,16$, and
complexity is optimally $O(N^2)$. Our straightforward multi-spin
implementation took roughly $3\times10^{-5}N^{2.5}$ seconds for  $10^6$ realizations with 64 samples each.
Figure \ref{fig:6} displays the number of samples used for the two
scenarios. Total single-core computation time  is more than 5 years for ASM and less than a year for Aztec. In actual fact, up to 100 CPU cores worked in parallel.     
%%%%%%%%%%%%%%%%%%%%%%%%% 
\begin{figure}[!t]
  \centering
  \includegraphics[width=0.85\linewidth]{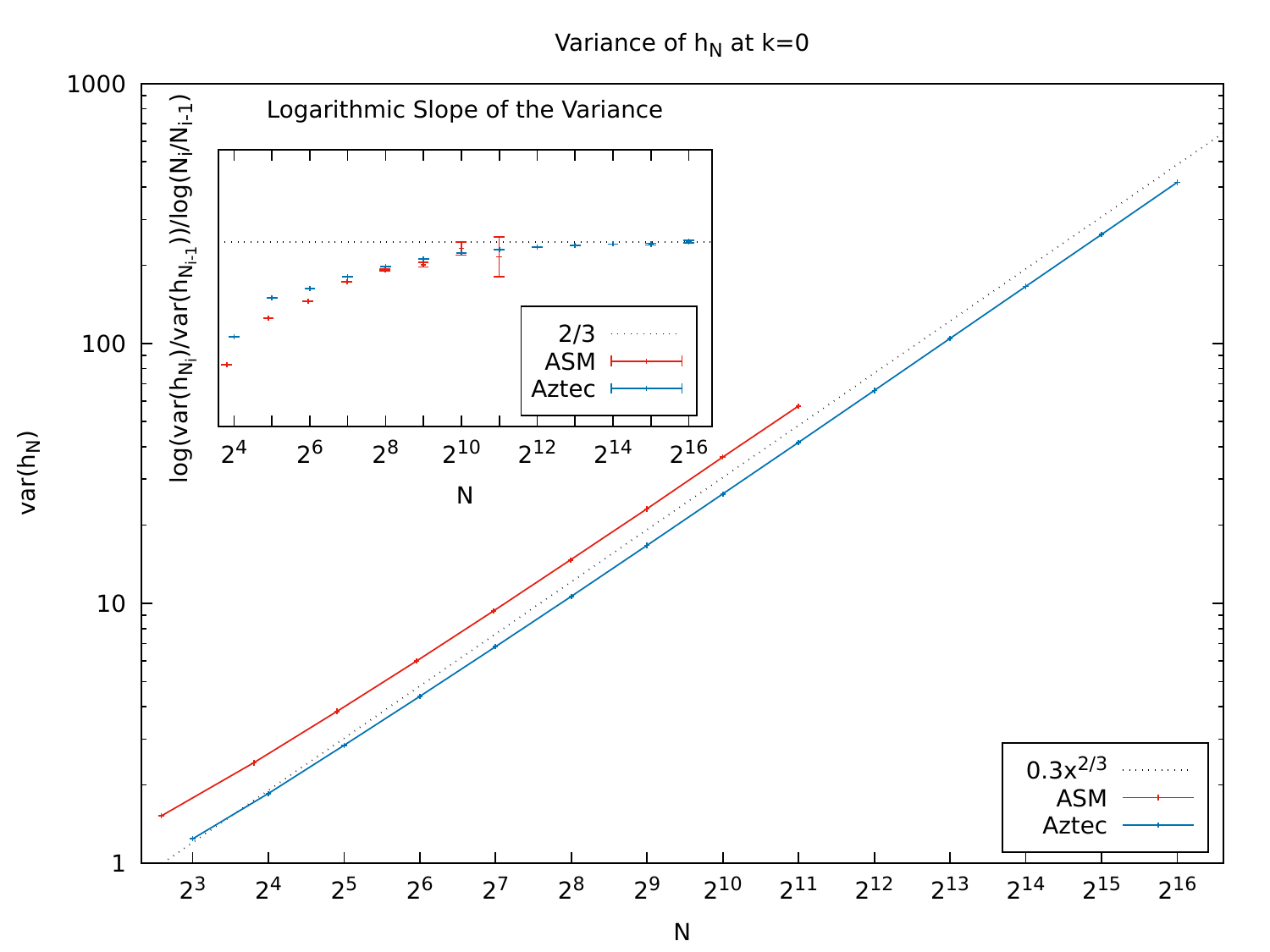}
  \caption{Dynamical scaling of the variance.}
  \label{fig:7}
\end{figure}
%%%%%%%%%%%%%%%%%%%%%%%%%%%%%

As a first test, the dynamical exponent $1/3$,  theoretically predicted in \eqref{3.2}, is
determined numerically from the variance of the edge function. 
For the Aztec diamond the corresponding logarithmic slope $2/3$ has been 
rigorously proven in \cite{J00}. For ASM the same exponent is confirmed in Figure \ref{fig:7} with high precision. 
\begin{figure}
  \centering
  \includegraphics[width=0.75\linewidth]{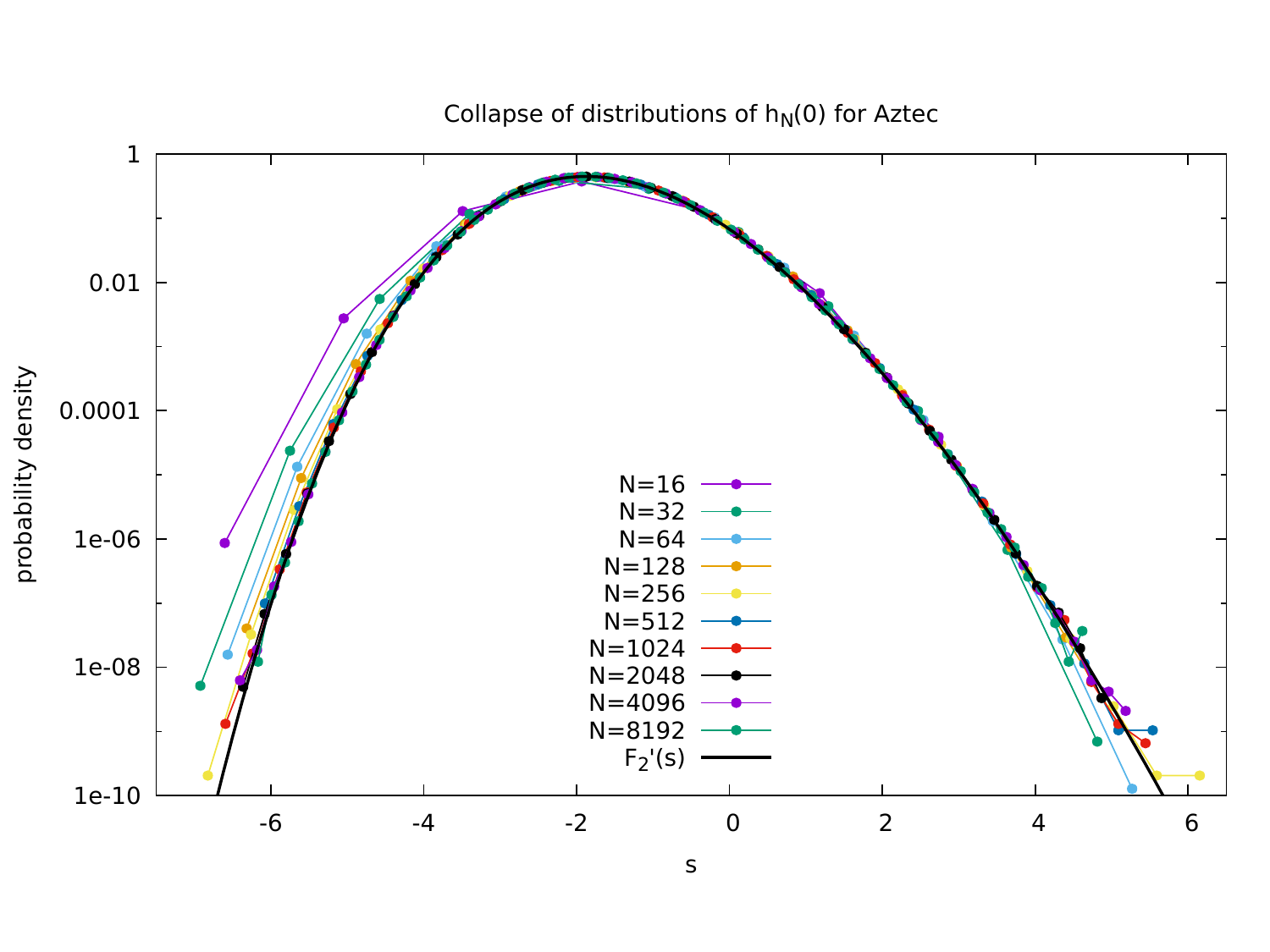}
  \vspace{-5ex}
  \caption{Data collapse of rescaled distributions for Aztec diamond.}
  \label{fig:8}
\end{figure}
%%%%%%%%%%%%%%%%%%%%%%%%%%%%%%%%%%%%%%%%

%%%%%%%%%%%%%%%%%%%%%%%%%%%%% 
\begin{figure}[!b]
  \centering
  \includegraphics[width=0.75\linewidth]{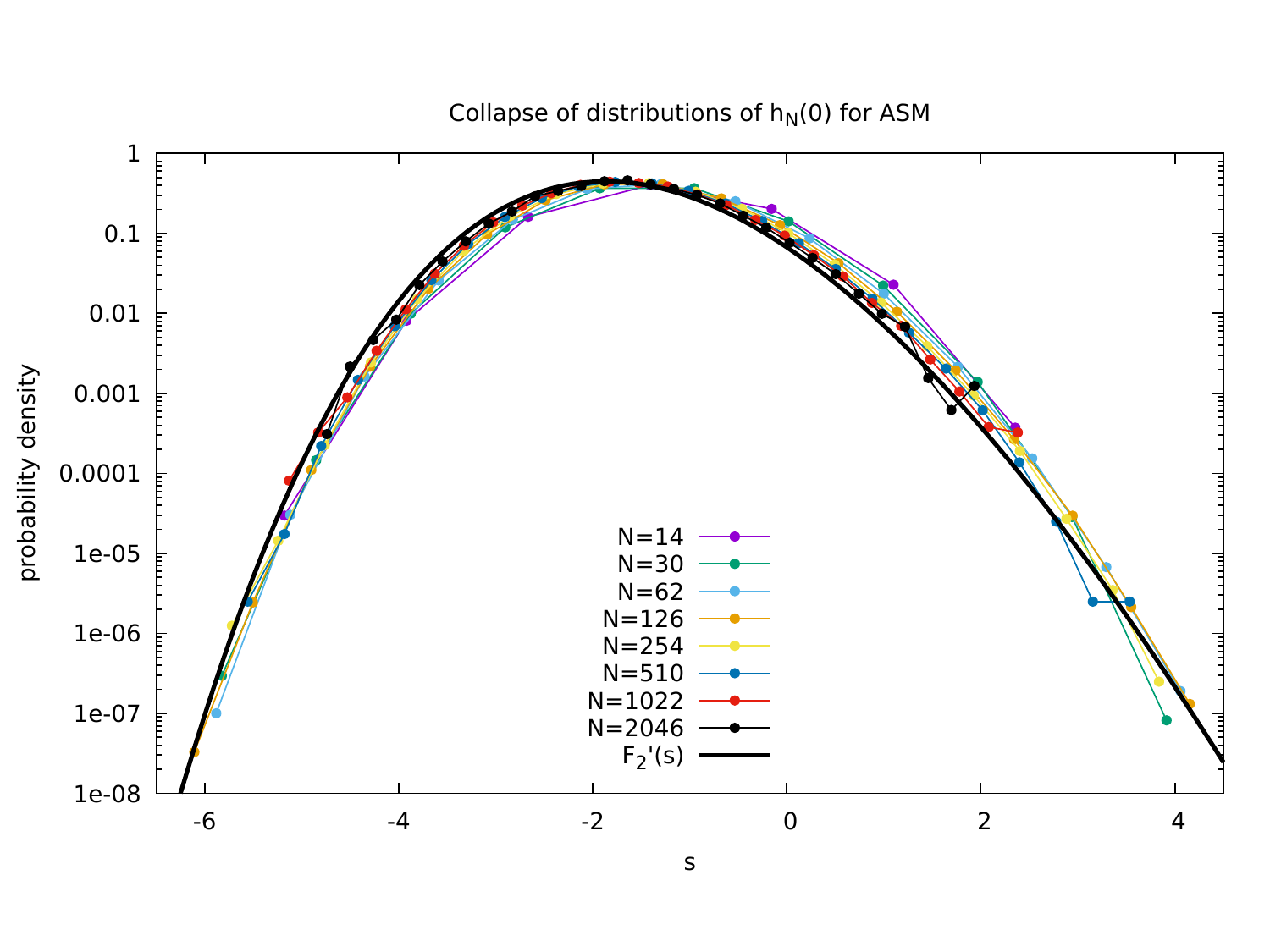}
  \vspace{-5ex}
  \caption{Data collapse of rescaled distributions for ASM.}
  \label{fig:9}
\end{figure}
%%%%%%%%%%%%%%%%%%%%%%%%%%%%%%%%%%

Using the known macroscopic circle shape $h_{\mathrm{ma}}$, first determined in \cite{JPS95}, together with the results from \eqref{3.5} and \eqref{3.12},
the model dependent parameters for the macroscopic shape and local
fluctuations for the Aztec point are obtained as
\begin{align}\label{aztecscaling}
  h^{\mathrm{Aztec}}_{\mathrm{ma}}(x)=1-\sqrt{\tfrac12-x^2},\qquad \Gamma^{\mathrm{Aztec}}(x)=\frac{(1-4x^2)^2}{2^{5/2}(1-2x^2)^{3/2}}\,,\qquad|x|<\frac12.
\end{align} 
Figure \ref{fig:8} displays the MC result for the
probability distributions of the random variable $h_N(0)$ in the Aztec case for different values of $N$, transformed to the KPZ scale
\begin{equation}
s=-\frac{h_N(0)-N h^{\mathrm{Aztec}}_{\mathrm{ma}}(0)} 
  {(\Gamma^{\mathrm{Aztec}}(0)N)^{1/3}}
\end{equation}
and compared to the limiting
GUE Tracy-Widom distribution $F_2'(s)$ \cite{TW93}.

At the ASM point the macroscopic shape determined in \cite{CP10} and the fluctuations derived with  \eqref{3.5} and \eqref{3.12} are
\begin{align}\label{asmscaling}
  h^{\mathrm{ASM}}_{\mathrm{ma}}(x)=2-\sqrt{3(1-x^2)},\qquad \Gamma^{\mathrm{ASM}}(x)=\frac{(1-4x^2)^2}{2\sqrt3(1-x^2)^{1/2}}\,,\qquad|x|<\frac12.
\end{align}
Figure \ref{fig:9} shows the collapse of probability distributions for $h_N(0)$ in the ASM case, this time on the scale
\begin{equation}
s=-\frac{h_N(0)-N h^{\mathrm{ASM}}_{\mathrm{ma}}(0)}{(\Gamma^{\mathrm{ASM}}(0)N)^{1/3}},
\end{equation}

%\newpage
 as predicted by KPZ scaling theory, and again compared to the
Tracy-Widom density $F_2'(s)$. Note that in the plots of Figure
\ref{fig:8} and \ref{fig:9} there is no free fitting parameter.

So far we took into account only the fluctuations of the facet edge
along the diagonal. More generally,  in the limit of large $N$, the scaling theory predicts for each tile
the probability to be in the frozen region of
the lower right corner, compare  with Figure
\ref{fig:1}. One expects the deviation from the properly rescaled
distribution function $F_2$ to be small,
\begin{equation}
  \label{eq:26}
  \mathrm{Prob}\big(h_N(k)<h\big)
  - F_2\Big(-\frac{h-N h_{\mathrm{ma}}(\tfrac{k}N)}
  {(\Gamma\big(\tfrac{k}N\big)N)^{1/3}}\Big)\simeq0.
\end{equation}
%%%%%%%%%%%%%%%%%%%%%%%%%%%%%%%%%
\begin{figure}[!t]
  \centering
  \includegraphics[width=.6\linewidth]{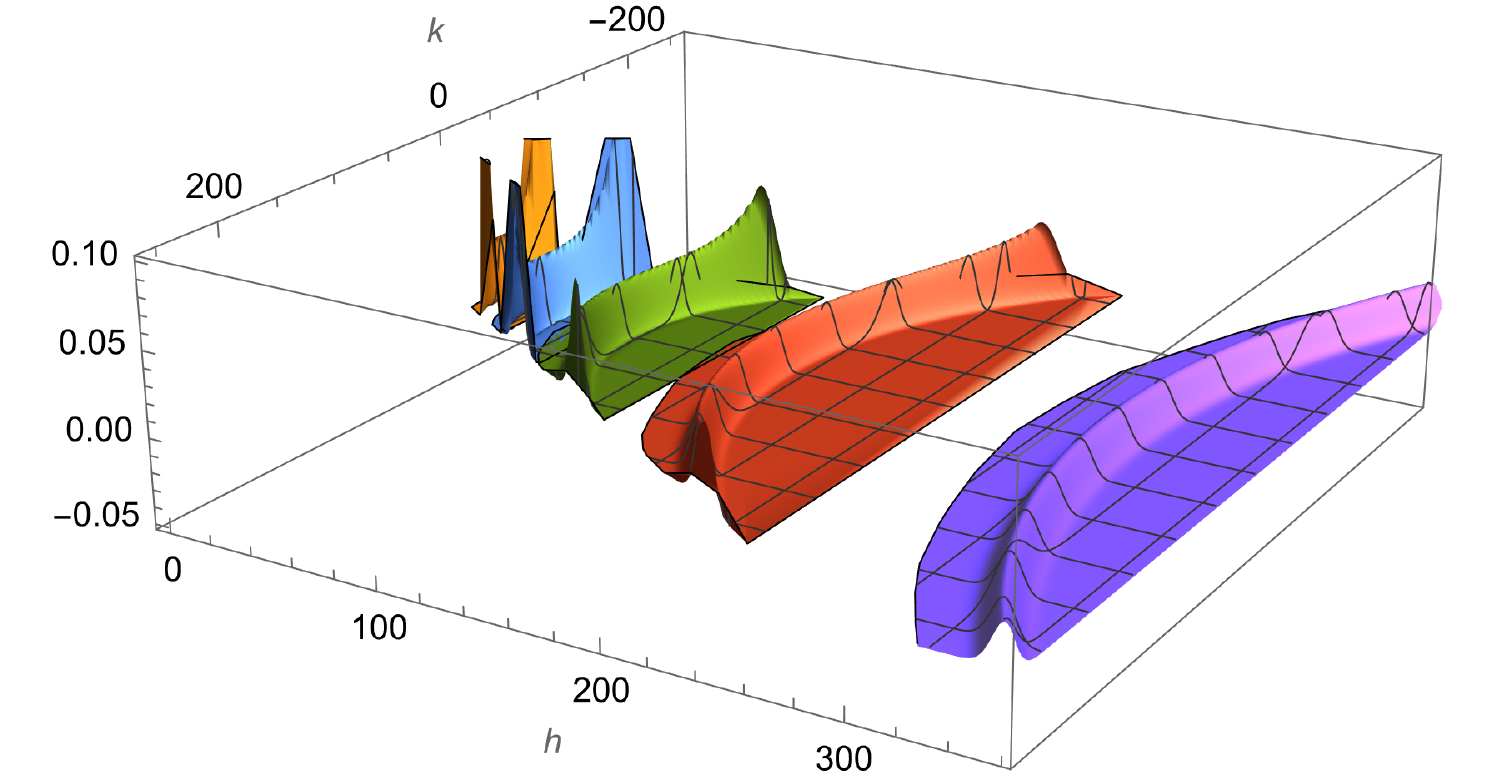}
  \caption{Deviations from the limiting Tracy-Widom distribution  at system sizes $N=64$ (yellow), $N=128$ (blue), $N=256$
  (green), $N=512$ (red) , and $N=1024$ (violet) for Aztec. The spikes reflect slow convergence for $|k|$ close to $\frac{N}2$.}
  \label{fig:10}
\end{figure}
%%%%%%%%%%%%%%%%%%%%%%%%%%%%%%%
\begin{figure}[!b]
  \centering
  \includegraphics[width=.6\linewidth]{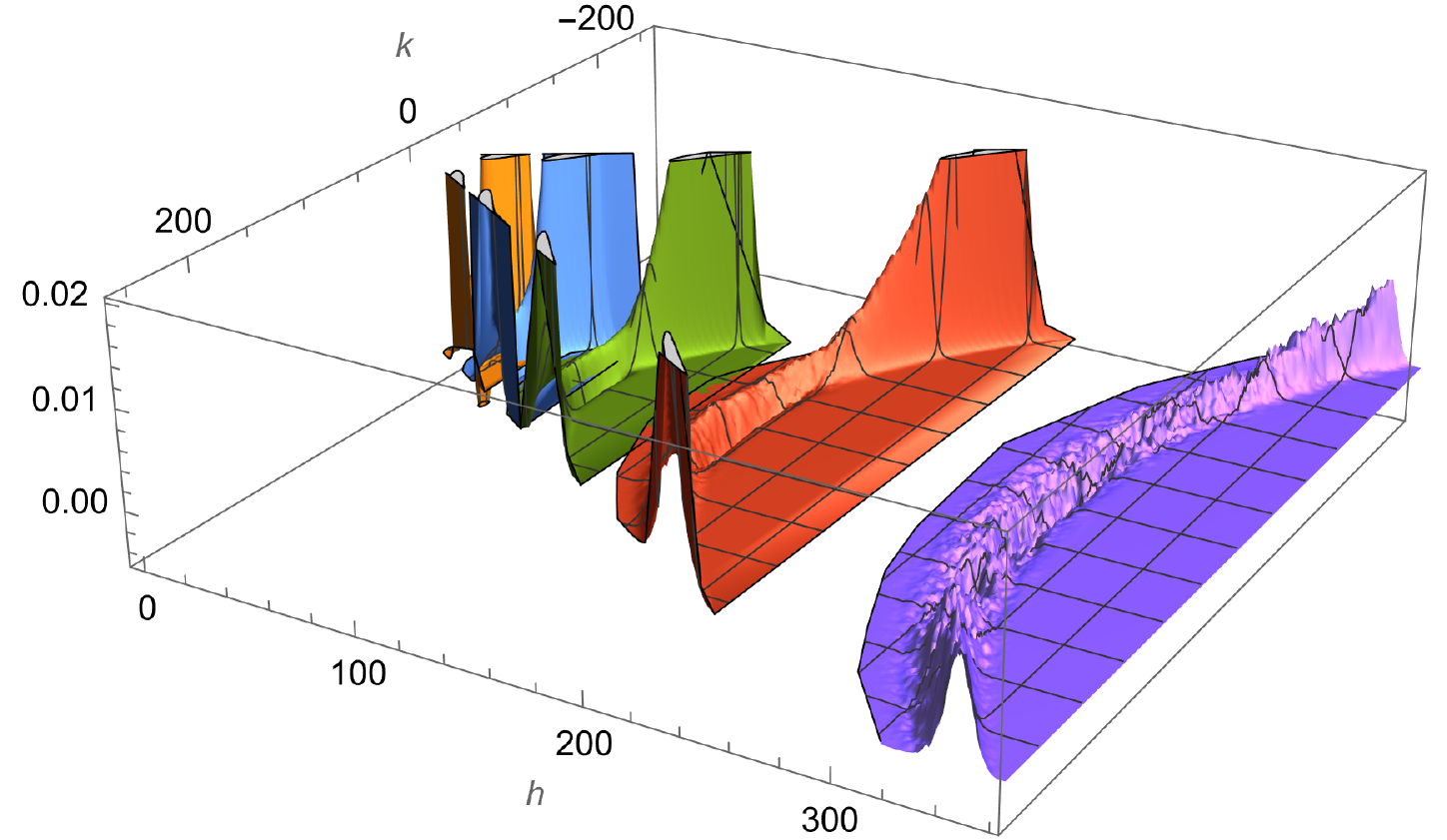}
  \caption{Deviations from the limiting Tracy-Widom distribution at
  system sizes $N=62$ (yellow), $N=126$ (blue), $N=254$
  (green), $N=510$ (red), and $N=1022$ (violet) for  ASM. The spikes
  for $|k|$ close to $\frac{N}2$ are much more pronounced than in the
  Aztec case.}
  \label{fig:11}
\end{figure}
%%%%%%%%%%%%%%%%%%%%%%%%%%%%%%%%%%%%%%%%%%%%%%%%%%
In Figures \ref{fig:10} and \ref{fig:11} we present for Aztec and ASM, respectively, the numerical value of the difference in \eqref{eq:26} as a function of $k$ and $h$. For various values of
$N$ plotted are the results in the range $|k|\leq 0.9 h$. Note that
for the extreme values of $k$, corresponding to $|\frac{k}{N}|$ close
to $\frac12$, the convergence to zero is particularly slow. For Aztec
the convergence to $0$ as $N\to\infty$ has been established. The error is expected to be of order
$N^{-1/3}$. Although the qualitative plot in Figure \ref{fig:11} is less
conclusive, we still conjecture a similar behaviour for the case of ASM.

Fixing  the origin of $k, h,N$ with  precision
higher than order $N^{1/3}$
is somewhat arbitrary for $h$ and $N$, while the origin of $k$ can be determined
by symmetry. We hope that by appropriately choosing these offsets or a more complicated 
correction term of order $1$ (size of a tile) the  
convergence rate can be improved. As known from other models
\cite{FF11,B23} convergence might be even of order $N^{-2/3}$.

\vspace*{6pt}%\\
%\newpage
\noindent
\textbf{Acknowledgements}.
We thank Alexander Hartmann for insightful reading of the manu\-script and an anonymous referee for his instructive comments. We gratefully acknowledge the Leibniz Supercomputing Centre for funding this project by providing computing time on its Linux-Cluster.\smallskip\\
\textbf{Data Availability}.
A synopsis of the Monte Carlo results for ASM and Aztec
is available at \texttt{https://www-m5.ma.tum.de/KPZ}.

\end{document}